\documentclass[12pt]{article}

\usepackage{amsmath}
\usepackage{amssymb}
\usepackage{mathtext}
\usepackage{graphicx}
\usepackage{psfrag}

\begin{document}

\title{On the motion of a heavy rigid body in an ideal fluid with circulation}
\author{Alexey V. Borisov, Ivan S. Mamaev\\
Institute of Computer Science, Udmurt State University\\
1, Universitetskaya str., 426034 Izhevsk, Russia\\
Phone/Fax: +7-3412-500295, E-mail: borisov@rcd.ru, mamaev@rcd.ru}
\maketitle

\begin{abstract}
Chaplygin's equations describing the planar motion of a rigid body in an
unbounded volume of an ideal fluid involved in a circular flow around the
body are considered. Hamiltonian structures, new integrable cases, and
partial solutions are revealed, and their stability is examined. The
problems of non-integrability of the equations of motion because of a
chaotic behavior of the system are discussed.
\end{abstract}

\newpage
\section{Introduction and a review of the known results.}
S.\,A.\,Chaplygin \cite{bor01} considered a general problem regarding the
forces and momenta that impact an arbitrary rigid body involved in a
parallel-plane motion in an unbounded volume of an ideal incompressible
fluid. More specific formulations were studied earlier by N.\,E.\,
Zhukovski \cite{bor02, bor03}, who considered the application of his
formula for the lifting force to the description of heavy body falling in
a fluid. However, these formulations yielded an unrealistic conclusion
that the rotational and translational motions are independent, thus
showing that additional analysis is required. The necessary analysis was
made by S.\,A.\,Chaplygin.

Chaplygin made general assumptions that the fluid motion is vortex-free
with a zero velocity at the infinity, and the fluid circulation around the
body is constant. He observed that this form of equations holds true if
the propeller propulsion does not change its direction with respect to the
body (aircraft) and the drag equilibrates this propulsion at any moment.
Although this assumption resulted in the conclusion that the aircraft will
have a directional stability, which is not true (the aircraft trajectory
may be winding), some general conclusions that can be drawn from this
study allow a remarkable mechanical interpretation. They are of both
theoretical and practical interest even in the context of modern applied
aerohydromechanics.

S.\,A.\,Chapygin \cite{bor04} considered the case of a circulation-free
planar flow around a body and suggested a remarkable form of equations,
which unfortunately he has not studied at all (equations with a similar
form describe the falling of a body of rotation in a fluid under the same
conditions). Originally written in 1890 as his student essay, this work was published
only in 1933 as part of Chaplygin's Collected Works.
  Independently and at about
the same time D.\,N.\,Goryachev \cite{new006} (1893) and V.\,A.\,Steklov \cite{bor05} (1894) obtained the same
equations and described their simplest qualitative properties. In particular,
V.\,A.\,Steklov showed that, as the body falls, the amplitude of its oscillations
with respect to the horizontal axis decreases while the frequency of these
oscillations increases. He made this remark in an addendum to his
book \cite{bor05}, in which the analysis of the asymptotic behavior was
made with some mistakes. In essence, V.\,A.\,Steklov formulated the problem of
asymptotic description of the behavior of a heavy rigid body during its fall. This
problem was solved by V.\,V.\,Kozlov~\cite{bor06}, who showed
that, under almost all initial conditions, the body tends to fall at a
uniform acceleration with its wider side up and oscillates around the
horizontal axis with an increasing frequency of the order of $t$ and a
decreasing amplitude with an order of $1/\sqrt{t}$. Asymptotic motions
with different numbers of half-turns were analyzed numerically
in~\cite{bor07}. The asymptotic behavior of a body falling without initial
impact was studied in \cite{new037}.\looseness=1

The effect of an abrupt ascent was described and studied in \cite{bor08}.
Under conditions of vortex-free flow around the body, it is assumed that,
in the initial moment, the wider side of the body is horizontal and the
body has a horizontal velocity. In the following moments, the body starts
moving downward. However, if its apparent mass in the lateral direction is
sufficiently large, the body will next abruptly move upward with its
narrower side up and rise higher than its initial elevation.

More general equations describing a non-planar motion of a heavy rigid body
in an unbounded volume of an ideal fluid involved in a vortex-free motion
and resting at the infinity were also obtained in~\cite{bor04} (more compact
form of these equations is presented in~\cite{bor09}). They
generalize the well-known Kirchhoff equations, which are known to neglect
the force of gravity. As was shown in \cite{bor09}, if a rigid body has
three mutually orthogonal planes of symmetry, this body,  when falling
freely, asymptotically tends to take the position when its axis with the
maximum added mass is vertical. Such body can also rotate around this
axis.

A particular case of existence of the Hessian integral for the general
equations~\cite{bor04} is described in \cite{bor10}. In this case, the
equations can be reduced to a simpler form analogous to the planar case.
It is worth mentioning that a planar motion of a rigid body in a resisting
medium is considered in \cite{bor10_2}. For various models of hydrodynamic forces,
some numerical and experimental  results concerned with the falling motion of a heavy body
in a fluid are presented in the papers \cite{new015, new016, new032, new035, new036}. An elementary analysis
of the motion of a body in a resisting medium was
first performed by Maxwell in \cite{new033}.

Now let us return to the problem of a planar circular motion of a rigid
body, which is the focus of this study. The equations of motion for this
problem were suggested by S.\,A.\,Chaplygin in~\cite{bor01}. It was
studied in \cite{bor11} with some specific assumptions (as compared with
\cite{bor01}).

Circulation makes the motion of the rigid body more complicated. In
\cite{bor11}, cases of stationary motion are found and their stability is
examined. An integrable case and certain classes of particular solutions
analogous to Zhukovski's solutions \cite{bor02,bor03} are proposed.

In this paper, we will propose the Hamiltonian form of the general
equations, and show new cases of integrability and classes of new
particular solutions. We will also show that in the general case, the
equations \cite{bor01} are not integrable and their behavior is chaotic.

\section{General equations of motion. Lagrangian and Hamiltonian descriptions.}
% Figure 1{\wfig<bb=0 0 55.3mm 51.9mm>{01.pcx}}

We choose a moving coordinate system $O_1 x_1 x_2$ which is fixed to the body. The
position of this system with respect to the fixed frame is characterized by
the coordinates $(x,y)$ of its origin $O_1$ and the rotation angle
$\varphi$. Let us assume that $(\xi,\eta)$ are the Cartesian
coordinates of the center of masses in the moving coordinates, $I$ is the
constant circulation around the body, $\rho$ is the fluid density; $v_1$,
$v_2$ are the projections of the linear velocity of the center $O_1$ onto
the moving axes. Let us assume also that $Q_1$, $Q_2$ are the projections of
the external forces onto the moving axes, $\omega$ is the angular velocity
of the body, and $M_1$ is the momentum of the external forces relative to
the center of masses.

Explicit evaluation of the forces and momenta associated with the
circulation flow around a body was used in \cite{bor01} to derive the
following equations, which are similar to the Kirchhoff
equations:
\begin{equation}
\begin{split}
\label{ur1}
a_1 \dot {v_1} & = a_1 \omega v_2 - \lambda v_2 - \zeta \omega -
Q_1,\\
a_2 \dot{v_2} & = -a_2 \omega v_1 + \lambda v_1 + \chi \omega - Q_2,\\
b \dot \omega & = (a_1 - a_2) v_1 v_2 + \zeta v_1 - \chi v_2 - M_1 + \xi Q_2
- \eta Q_1,\\
\dot \varphi & =  \omega,\\
\dot x & = (\boldsymbol{\alpha}, \boldsymbol{v}) = v_1 \cos \varphi - v_2 \sin \varphi,\\
\dot y & = (\boldsymbol {\gamma}, \boldsymbol{v}) = v_1 \sin \varphi + v_2 \cos \varphi,
\end{split}
\end{equation}

where $\boldsymbol{\alpha} = (\cos \varphi - \sin \varphi)$, $\boldsymbol{\gamma} = (\sin \varphi, \cos
\varphi)$, $\boldsymbol{v} = (v_1, v_2)$.

In equations \eqref{ur1} $a_1, a_2, b$ are added masses and momentum, $\lambda=I\rho$,
$\zeta=I\rho\mu_1$, $\chi=I\rho\lambda_1$.
Parameters $\zeta$, $\chi$, proportional to the circulation $I$, are
associated with the asymmetry of the body, and their evaluation is
described in \cite{bor01}. In the general case with circulation and asymmetry of the body
$\zeta \ne 0$, $\chi \ne 0$, these parameters cannot be eliminated by any
choice of a coordinate system fixed to the body.

{\textbf{Remark.}
\textit{Equations \eqref{ur1} were proposed by S.A.Chaplygin in 1926 and somewhat
later (and presumably independently) were derived by Lamb and Glowert in
1929 \cite{bor13}. Unlike Chaplygin~\cite{bor01}, the behavior of
solutions of these equations was not analyzed in these works.}

\textit{It is worth mentioning that equations similar to \eqref{ur1} describe the planar
motion of a rigid body in an ideal fluid with a uniform vorticity. The
motion of a circular cylinder in a fluid with a uniformly distributed
vorticity (for different boundary conditions) was analyzed by Prudeman and
Taylor \cite{bor13}. In this case, the equations of motion are reduced to
a simplest linear system. The general (even planar) motion of an arbitrary
rigid body is very complicated and has not been analyzed yet.}

If all forces are potential, we have
\begin{equation}
\begin{split}
\label{ur2}
Q_1 & = {\frac{\partial U}{\partial x}} \cos \varphi + {\frac{\partial U}{\partial y}} \sin \varphi,\\
Q_2 & = -{\frac{\partial U}{\partial x}} \sin \varphi + {\frac{\partial U}{\partial y}} \cos \varphi,\\
M_1 & - \zeta Q_2 + \eta Q_1 = {\frac{\partial U}{\partial {\varphi}}},
\end{split}
\end{equation}
where the function $U = U(x, y, \varphi)$ is the potential. In this
case, equations \eqref{ur3} have the integral of energy
\begin{equation}
\label{ur3}
H = T + U =
\frac{1}{2} (a_1 v_1^2 + a_2 v_2^2 + b \omega^2) + U,
\end{equation}
where the kinetic
energy $T$ of the (``body $+$ fluid'') system is written in the diagonal
form after translations and rotation of the moving coordinate system.

It can be shown that the forces caused by circulation are generalized
potential forces and the equations of motion (in the case of potential
external forces) can be written in the Lagrangian form
\begin{equation}
\begin{split}
\label{ur4}
\frac{d}{dt}\left({\frac{\partial L}{\partial v_1}}\right) &=
\omega{\frac{\partial L}{\partial v_2}} + {\frac{\partial L}{\partial x}}\alpha_1
+ {\frac{\partial L}{\partial y}} \gamma_1\\
\frac{d}{dt}\left({\frac{\partial L}{\partial v_2}}\right) &=
-\omega{\frac{\partial L}{\partial v_1}} + {\frac{\partial L}{\partial x}}\alpha_2
+ {\frac{\partial L}{\partial y}} \gamma_2\\
\frac{d}{dt}\left({\frac{\partial L}{\partial \omega}}\right) &=
v_2{\frac{\partial L}{\partial v_1}} - v_1{\frac{\partial L}{\partial v_2}} + {\frac{\partial L}{\partial {\varphi}}}\\
\dot \varphi &= \omega,\mbox{  }\dot x = (\boldsymbol{\alpha}, \boldsymbol{v}),\mbox{  } \dot y = (\boldsymbol{\gamma},
\boldsymbol{v}),
\end{split}
\end{equation}
where the Lagrangian function has the form
\begin{equation}
\label{ur5}
L = T - U - (x \sin \varphi - y \cos \varphi)\left(\frac{\lambda}{2}v_1 +
\chi\omega\right) - (x \cos \varphi + y \sin \varphi)\left(\frac{\lambda}{2}v_2 +
\zeta \omega\right).
\end{equation}

{\textbf{Remark.}
\textit{The Lagrangian function is chosen in a symmetric calibration; recall that
the summands that are linear with respect to velocities are determined to
a total differential of an arbitrary function.}

Equations \eqref{ur4} are Poincar\'{e} equations on the group of motions of
the plane $E(2)$~\cite{bor10}; using the Legendre transformation we can
represent them in the Hamiltonian form (Poincar\'{e}--Chetaev equations) with
a Hamiltonian containing terms linear in the momenta. It was found
\cite{bor10_2}, that the equations of motion for generalized potential
systems can be conveniently represented in the slightly modified
variables:
$$p_1 = {\frac{\partial T}{\partial v_1}} = a_1 v_1\mbox{, } p_2 = {\frac{\partial T}{\partial v_2}} = a_2
v_2\mbox{,  } M = {\frac{\partial T}{\partial {\omega}}}b \omega.$$
Using the Legendre transformation
for new variables, we find the Hamiltonian:
\begin{equation}
\label{ur6}
H = (\boldsymbol{p}, \boldsymbol{v}) - L = \frac{1}{2}\left(\frac{p_1^2}{a_1} +
\frac{p_2^2}{a_2} + \frac{M^2}{b}\right) + U(x,y,\varphi).
\end{equation}
The obtained equations of motion have the form similar to that of the
Poincar\'{e}--Chetaev equations:
\begin{equation}
\begin{split}
\label{ur7}
\dot p_1 &= p_2 {\frac{\partial H}{\partial M}} - {\frac{\partial H}{\partial x}}\alpha_1 - {\frac{\partial H}{\partial y}}\gamma_1 -
\lambda{\frac{\partial H}{\partial p_2}} - \zeta{\frac{\partial H}{\partial M}},\\
\dot p_2 &= -p_1{\frac{\partial H}{\partial M}} - {\frac{\partial H}{\partial x}}\alpha_2 - {\frac{\partial H}{\partial y}}\gamma_2 +
\lambda{\frac{\partial H}{\partial p_1}} - \chi{\frac{\partial H}{\partial M}},\\
\dot M &= p_1{\frac{\partial H}{\partial p_2}} - p_2{\frac{\partial H}{\partial p_1}} - {\frac{\partial H}{\partial {\varphi}}} + \zeta
{\frac{\partial H}{\partial p_1}} - \chi{\frac{\partial H}{\partial p_2}},\\
\dot \varphi &= {\frac{\partial H}{\partial M}}\mbox{, } \dot x = \left(\boldsymbol{\alpha}, {\frac{\partial H}{\partial {\boldsymbol{p}}}}\right)
\mbox{, } \dot y = \left(\boldsymbol{\gamma},{\frac{\partial H}{\partial {\boldsymbol{p}}}}\right)
\end{split}
\end{equation}
and the Poisson bracket of these variables contains
``additional circulation terms'' (hygroscopic terms):
\begin{equation}
\begin{split}
\label{ur8}
\{M_1, p_1\} = - p_2 + \zeta\mbox{,  } \{M, p_2\} = p_1 - \chi\mbox{,  } \{p_1,p_2\} = - \lambda,\\
\{M,\varphi\} = -1\mbox{,  }\{M,x\} = \{M,y\} = 0, \\
\{p_1,x\} = - \alpha_1\mbox{,  }\{p_2,x\} = - \alpha_2\mbox{,  } \{p_1,y\} = -
\gamma_1\mbox{,  }\{p_2,y\} = - \gamma_2.
\end{split}
\end{equation}

The rank of this Poisson structure is equal to 6; therefore, the
system~\eqref{ur7} can be reduced to a canonical system with three degrees
of freedom.

\section{Motion in the gravity field.}

In this case, the potential energy of the system can be written in the form
$U = \mu(y + \xi \sin \varphi + \eta \cos \varphi)$,
and its Hamiltonian \eqref{ur7} has the form
\begin{equation}
\label{ur9}
H = \frac{1}{2}\left(\frac{p_1^2}{a_1} + \frac{p_2^2}{a_2} +
\frac{M^2}{b}\right)+
\mu(y + \xi \sin \varphi + \eta \cos \varphi).
\end{equation}

The system \eqref{ur7}, \eqref{ur9} admits an autonomous and a
non-autonomous integral, which correspond to the projections of the
momentum on the fixed axes $Ox$ and $Oy$ (see figure~1):
\begin{equation}
\begin{split}
\label{ur10}
p_x & = p_1 \cos \varphi - p_2 \sin \varphi + \lambda y + \zeta
\sin \varphi
- \chi \cos \varphi = c_1\mbox,\\
p_y^{} & = p_2 \sin \varphi + p_2 \cos \varphi - \lambda x - \zeta \cos
\varphi - \lambda \sin \varphi = \mu t + c_2,\\
c_1 & =const\mbox{,  } c_2 = const,
\end{split}
\end{equation}
which are not commutative $\{p_x, p_y\} = \lambda$.

To perform a reduction by one degree of freedom we express $y$ from $p_x=c_1$ and
substitute it into the Hamiltonian ~\eqref{ur9}:
\begin{equation}
\begin{split}
\label{ur11}
H = \frac{1}{2}\left(\frac{p_1^2}{a_1} + \frac{p_2^2}{a^2} +
\frac{M^2}{b}\right)+ \frac{\mu}{\lambda}\left(-p_1 \cos \varphi
+ p_2 \sin \varphi\right) + \\
+\mu \left(\left(\xi - \frac{\zeta}{\lambda}\right) \sin \varphi +
\left(\eta + \frac{\chi}{\lambda}\right)\cos \varphi\right).
\end{split}
\end{equation}

This reduced Hamiltonian depends only on the variables $p_1$, $p_2$, $M$
and $\varphi$, the Poisson brackets of which, according to \eqref{ur8},
form a closed subalgebra with a rank of 4. Thus, we have a reduced system
with two degrees of freedom, which can be written in the canonical form
(see below). However, we will not use the canonical form, but algebrise the
reduced system to even greater extent with the help of dependent variables
$\gamma_1 = \sin \varphi$, $\gamma_2 = \cos \varphi:$
\begin{equation}
\begin{split}
\label{ur12}
H & = \frac{1}{2}\left(\frac{p_1^2}{a_1} + \frac{p_2^2}{a_2} +
\frac{M^2}{b}\right)- \frac{\mu}{\lambda} \left((\boldsymbol{p} - \boldsymbol{s}), {\mathbf {J}}\boldsymbol{\gamma}\right) +
\mu(\boldsymbol{r}, \boldsymbol{\gamma}),\\
\dot {\boldsymbol{p}} & = {\frac{\partial H}{\partial {\boldsymbol{M}}}}{\mathbf{J}}(\boldsymbol{p}-\boldsymbol{s}) - \mu \boldsymbol{\gamma} - \lambda
{\mathbf{J}}{\frac{\partial H}{\partial {\boldsymbol{p}}}},\\
\dot {M} & = \left(\boldsymbol{p} - \boldsymbol{s}, {\mathbf{J}}{\frac{\partial H}{\partial {\boldsymbol{p}}}}\right) - \mu \left(\boldsymbol{r}, {\mathbf{J}}
\boldsymbol{\gamma}\right),\\
\dot {\boldsymbol{\gamma}} & = {\frac{\partial H}{\partial M}} {\mathbf{J}} \boldsymbol{\gamma},
\end{split}
\end{equation}
where $\boldsymbol{p} = (p_1, p_2)$, $\boldsymbol{\gamma} = (\gamma_1, \gamma_2)$, ${\mathbf{J}} =
\biggl\|
\begin{array}{cc}
0 & 1\\
-1 & 0
\end{array}
\biggr\|,
$ $\boldsymbol{s} = (\chi,\zeta)$, $\boldsymbol{r} = (\xi, \eta)$.

The system has two obvious integrals~--- energy $H$ and geometric integral
$\gamma_1^2 + \gamma_2^2 = 1$. Since the system is Hamiltonian, for being
integrable it requires one more additional integral (although this system can be
integrated by the Euler--Jacobi method because the equations \eqref{ur12}
conserve the standard invariant measure).

Note that the system \eqref{ur12} has an important mechanical meaning, its
form is as simple as that of the Euler--Poisson equations, and analogous
formulations of problems are quite meaningful for this system. The first
aspect relates to the integrability of the system \eqref{ur12}.

{\small {\bf Canonical variables.} Using the algorithm described in
\cite{bor10}, it is easy to construct canonical variables analogous to the Anduaye
variables in the rigid body dynamics. In conventional denotations,
we have
$$
M = L\mbox{,  } p_1 - \chi = \sqrt{2\lambda(G - L) \cos \varphi}\mbox{,  } p_2 - \zeta =
\sqrt{2\lambda(G - L)}\sin l,$$
$$M + \frac{(p_1 - \chi)^2 + (p_2 - \zeta)^2}{2\lambda} = G\mbox{,  } \varphi =
g,$$
where $l$, $L$ and $g$, $G$ are two pairs of canonical conjugate
variables $(\{l,L\} = \{g, G\} = 1)$. These variables can be useful for
construction of action-angle variables and for application of methods of
the Hamiltonian theory of perturbations.}

\section{Integrable cases.}

The following integrable cases are known.

{\bf 1.} $\mu = 0$ (S.\,A.\,Chaplygin, 1926~\cite{bor01})~---the
absence of the gravity field.

Additional quadratic integral has the form
\begin{equation}
\label{ur13}
F =\frac{1}{2}\left((p_1 - \chi)^2 + (p_2 - \zeta)^2\right) + \lambda M,
\end{equation}
and the system \eqref{ur12} can be reduced, similar to the classical Euler--Poinsot
case, to a system of three equations for
variables $M$, $p_1$, and $p_2$. Chaplygin in \cite{bor01} noted that explicit integration  involves
some very complicated quadrature, which is expressed in elliptic functions
under the condition $\chi = \zeta = 0$.

{\bf 2.} $\zeta = \chi = 0$, $a_1 = a_2 = 1$ (V.\,V.\,Kozlov, 1993
\cite{bor11})~--- a case of dynamical symmetry. This case is analogous to
the Lagrange case, although the additional integral is quadratic with
respect to momenta
\begin{equation}
\label{ur14}
F = \frac{1}{2} \frac{M^2}{b} + \mu (\boldsymbol{r}, \boldsymbol {\gamma}).
\end{equation}
Let us consider a new case of integrability.

{\bf 3.} Let us assume that $\chi = -\lambda \eta$, $\zeta = \lambda \xi$ (i.\,e.
$\boldsymbol{s} = - \lambda {\mathbf J} \boldsymbol{r}$). Additional quadratic integral has the form
\begin{equation}
\label{ur15}
F = \frac{1}{2} M^2 - \lambda b (M + (\boldsymbol p, {\mathbf{J}} \boldsymbol{r})) + \mu
b (\boldsymbol{r}, \boldsymbol{\gamma}).
\end{equation}

Except the cases mentioned above, this system doesn't have other (general)
cases with an additional integral linear or quadratic with respect to
phase variables. This simple statement can be proved by direct enumeration
of variants with the use of the method of undetermined coefficients. The
question of the existence of higher-order integrals is still an open
question.

Let us consider a linear invariant relation analogous to the Hess
case for the Euler--Poisson equations.

{\bf 4.} Let us suppose that $\zeta = \chi = 0$, $\eta = \pm \sqrt{b\left({a_2^{-1}} -
{a_1^{-1}}\right)}$. In this case,
\begin{equation}
\begin{split}
\label{ur16}
H & = \frac{1}{2}\left(\frac{p_1^2}{a_1} + \frac{p_2^2}{a_2} +
\frac{M^2}{b}\right)- \frac{\mu}{\lambda} \left((p_1 - \chi)\gamma_2 + p_2
\gamma_2\right)\pm \mu \sqrt{b\left({a_2^{-1}} -
{a_1}^{-1}\right)}\gamma_2,\\
F & = M \pm \sqrt{b\left({a_2^{-1}} - {a_1^{-1}}\right)}
\end{split}
\end{equation}
and at the level $F = 0$ we have $\dot F = 0$.

{\textbf{Remark.}
\textit{
Note that in the case with zero circulation $(\lambda = 0)$, the
reduction to the system~\eqref{ur12} is impossible since $\frac{\mu}{\lambda}
= \infty$. Nevertheless, integrals $p_x = \pi_1$, $p_y = \pi_2 + \mu t$,
$\pi_1,\pi_2 = const$ allow us to eliminate $p_1$, $p_2$ from the
equations (rather than from the Hamiltonian) and obtain, similar to
Chaplygin \cite{bor04}, the non-autonomous ``pendulum''-type equation for
$\ddot \varphi$}:
\begin{equation}
\begin{split}
\label{ur17}
\ddot{\varphi} = \frac{\mu^2 t^2}{2b} (a_2^{-1} - a_1^{-1})\sin 2 \varphi
 + \frac{\mu t}{b}\biggl((a^{-1}_2 - a_1^{-1})(c_1^{} \cos 2\varphi + c_2 \sin 2
\varphi)-\frac{\chi \cos \varphi}{a_1} + \frac{\zeta \sin \varphi}{a_2}\biggr)
-\\
- \frac{\mu}{b}(\xi\cos \varphi - \eta \cos \varphi)
+ \frac{(a_1^{-1} - a_2^{-1})}{2b}\left((c_1^2 - c_2^2) \sin 2 \varphi -
2 c_1^{} c_2^{} \cos 2 \varphi\right)-\\
- \frac{\mu}{b}(\xi \cos \varphi - \eta \sin \varphi) +
\frac{\pi_1}{b}\left(\frac{\chi \sin \varphi}{a_1} +
\frac{\zeta \cos \varphi}{a_2}\right) - \\
-\frac{\pi_2}{b}\left(\frac{\chi \cos \varphi}{a_1} -
\frac{\zeta \sin \varphi}{a_2}\right)\
\end{split}
\end{equation}
\textit{which in particular cases was examined in~\cite{bor06,bor07,bor08}}.

\section{Chaplygin's case. Bifurcation analysis.}

Let us consider the system \eqref{ur9} at $\mu=0$, $\rho =\chi =0$. As was
noted above, equations for $M$, $p_1$, $p_2$ can be separated and have the
form
\begin{equation}
\label{ur18}
\dot p_1^{} = p_2 M - \lambda \frac{p_2}{a_2}\mbox{,  }
\dot p_2^{} = -\frac{p_1 M}{b} + \lambda \frac{p_1}{a_2}\mbox{,  }
\dot M = \left(\frac{1}{a_2} - \frac{1}{a_1}\right)p_1 p_2.
\end{equation}
The common level of the first integrals
\begin{equation}
\label{ur19}
H = \frac{1}{2}\left(\frac{p_1^2}{a_1} + \frac{p_2^2}{a_2} +
\frac{M^2}{b}\right)= h = const\mbox{,  }
F = M + \frac{p_1^2 + p_2^2}{2 \lambda} = c = const
\end{equation}
is represented by closed curves formed at the
intersection of an ellipsoid and an elliptic paraboloid; these curves are
analogous to centroid lines in the classical Euler--Poinsot problem
(figure~2). It is easy to find and describe the particular solutions of
\eqref{ur18} which are analogous to permanent rotations in the latter problem.
%figure 2. {03.eps}

For the sake of definiteness, let us assume that $\lambda > 0$ and $a_1 >
a_2$.

I. $p_1 = p_2 =0$, $M = c = const$, $h = \frac{1}{2b}c^2$. In this case,
the body rotates uniformly $(\dot \varphi = \frac{c}{b} = const)$ around
the origin of the moving coordinate system $O_1$ (see figure~1), though
this origin is fixed (and in the general case does not coincide with the
center of mass). This solution is unstable when $\frac{\lambda b}{a_1^{}} < c
< \frac{\lambda b}{a_2}$ and stable otherwise (the stability on the
boundaries requires special analysis).

II, III. There exists a pair of analogous solutions of the
form
\begin{equation}
\label{ur20}
p_i^{} = 0\mbox{,  } M = \frac{\lambda}{a_j}b = const\mbox{,  }
p_j^2 = 2 \lambda \left(c - \frac{b \lambda}{a_j}\right),
\end{equation}
where in one case, $i=1$, $j=2$, while in the other case, $i=2$, $j =1$.
Each of these solutions exists under condition $c > \frac{b\lambda}{a_j}$, and
the constants of the integrals for them are related by
\begin{equation}
\label{ur21}
 h = c
\frac{\lambda}{a_j} - b \frac{\lambda^2}{a_j^2}.
\end{equation}

%figure 3. {02.eps}
In the case of such motion, the solid body uniformly rotates around the
origin of a fixed coordinate system, and the principal axes of the body
always pass through this fixed center. The body is directed toward the
center of rotation by its wider side in one motion and narrower side in
the other. The solution $p_2 = 0$, $h = \frac{\lambda}{a_1}c -
\frac{1}{2}\frac{\lambda^2}{a_1^2}b$ is always stable while the solution $p_1
= 0$, $h = \frac{\lambda}{a_2}c - \frac{1}{2}\frac{\lambda^2}{a_2^2}b$ is always
unstable. Thus, stable motions are such that the body is directed toward
the fixed point by its narrower side.

The bifurcation diagram is shown in figure 3. The straight lines represent
the particular solutions II and III; they are tangent to the parabola
corresponding to solution I in the points where admissible values of the
integral $c$ begin. The domain of possible motions on the bifurcation
diagram is hatched.

The bifurcation diagram presents three different intervals of the values
of energy~$h$
$$\left(0, \frac{b\lambda^2}{2 a_1^2}\right)\mbox{,  }
\left(\frac{b\lambda^2}{2 a_1^2}, \frac{b\lambda^2}{2 a_2^2}\right)\mbox{,  }
\left(\frac{b\lambda^2}{2 a_2^2}, \infty \right),$$ for which the patterns of the
trajectories corresponding to different values of $c$ are qualitatively
similar (figure 2).

Let us consider explicit formulas for bi-asymptotic motions for solution
I, which is unstable at $\frac{\lambda b}{a_1} < c < \frac{\lambda b}{a_2}$. These
solutions are homoclinic (see figure 2\,{\it b}) and have the form
\begin{equation}
\begin{split}
\label{ur22}
M & = c - \frac{2 B d_1^{} d_2^{}}{\cosh 2 \tau - d_1^2 + d_2^2},\\
p_1 & = \pm \sqrt{\frac{8 \lambda B}{d_1^2}}\frac{d_1^2 d_2^2 \cosh \tau}
{\cosh 2 \tau - d_1^2 + d_2^2}\mbox{,  }
p_2 = \sqrt{\frac{8 \lambda B}{d_2^2}}\frac{d_1^2 d_2^2 \sinh \tau}{\cosh 2 \tau - d_1^2 +
d_2^2},\\
\varphi & = \frac{c}{b}(t - t_0^{}) - 2 \arctan\left(\frac{1+ d_1^2 - d_2^2}{2 d_1 d_2}\tanh
\tau\right)\mbox{,  } \tau = \frac{\sqrt{b_1b_2}}{2b}t,
\end{split}
\end{equation}
where $d_1^2 = \frac{b_1}{B}$, $d_2^2 = \frac{b_2}{B}$, $b_1 = 2(c -
\lambda
b a_1^{-1})$, $b_2 = 2(\lambda b a_2^{-1} - c)$, $B = b_1 + b_2$, and
different signs correspond to the two different separatrices.

\section{Perturbation of Chaplygin's case. Splitting of separatrices.}

Let us consider a ``perturbed'' Hamiltonian \eqref{ur9}, $H = H_0 + \mu
H_1$ with $\mu$ regarded as a small parameter. Here, $H_0$ is the
Hamiltonian of the integrable Chaplygin problem, $H_1 = U$ is the
perturbation function. One of the dynamic effects that prevent the
existence of an additional analytical integral for the perturbed system is
the splitting of separatrices, which are coupled in the case of the
unperturbed system. To the first order of the theory of perturbations at
small $\mu$, the splitting of separatrices is determined by the value of
the Poincar\'{e}--Mel'nikov integral \cite{bor12}
\begin{equation}
\label{ur23}
J =\int_{-\infty}^{\infty} \{F_0, H_1\}dt,
\end{equation}
calculated along asymptotic solutions of the unperturbed problem. Here
$F_0$ is an integral of the unperturbed system.

If the integral (as a function of the parameter on the separatrix) has a
simple zero, then separatrices split and transversally intersect. In the case of two degrees
of freedom this makes the perturbed problem non-integrable \cite{bor12}.

We take the integral~\eqref{ur19}~$F = M^2 + \frac{p_1^2 + p_2^2}{2 \lambda}$ as
$F_0$.

On homoclinic asymptotic solutions \eqref{ur22}, the integral \eqref{ur23}
can be explicitly evaluated using residues:
\begin{equation}
\begin{split}
\label{ur24}
J(g_0) & = J_Q + J_R,\\
J_Q & = - 2\pi \frac{\sqrt{2}\mu b}{\sqrt{\lambda B}} \frac{e^{\frac{\pi S}{2}}\sinh \left(\frac{\pi}{2} - \alpha\right)S}{\sinh \pi
S}\sin g_0,\\
J_R & = - 4 \pi \frac{\mu b}{B} \frac{e^{\frac{\pi S}{2}}\cosh \left(\frac{\pi}{2} - \alpha\right)S}{\sinh \pi S}
\left(\xi \cos g_0 - \eta \sin g_0\right),
\end{split}
\end{equation}
where $S = 2c(b_1b_2)^{-1/2}$, $\cos \alpha  = d_1$, $\sin \alpha =d_2$.

The integral \eqref{ur24} has a simple zero, except for the special case
of $B=0$. This allows us to conclude that the perturbed system \eqref{ur9}
is non-integrable, except for the case of $a_1^{} = a_2^{}$. Note that it
is under this condition that the integrals~\eqref{ur14}, \eqref{ur15}
were obtained. This condition is necessary but not sufficient for the
integrability.

The chaotic behavior of the system \eqref{ur9}, associated with the
splitting of separatrices and non-integrability, is illustrated in
figure~4, which represents a Poincar\'{e} section of the phase flow of the
system \eqref{ur12} on the energy level of $H = const$. The figure shows
the behavior of the split separatrices and the stochastic layer that forms
near them. The figure also presents the projection of the section at the
energy level of $H = h = const$, which is determined by the relationship
$\varphi = \pi = const$, on the plane $(p_1,p_2)$. As before, the
parameter values are taken as follows: $\lambda = 1$, $b = 2$, $a_1 = 2 a_2 =
1$, $\chi = \zeta = 0$ and for the energy level at these parameter values
we assume that $h=2$.

Let us discuss the issue of the possible fall of a body in the presence of circulation.
Recall that the body moving in the absence of circulation asymptotically tends to fall
with its wider side down \cite{bor06}.

{\it If $\lambda\neq 0$, the moving body will stay in a finite-width band parallel to the $x$ axis.}

Indeed, according to \eqref{ur10}, we have
\begin{equation}
\begin{split}
\label{ur25}
x&=-\frac\mu\lambda
t+\lambda^{-1}\left((p_1-\chi)\sin\varphi+(p_2-\zeta)\cos\varphi-c_2\right),\\
y&=-\lambda^{-1}\left((p_1-\chi)\cos\varphi-(p_2-\zeta)\sin\varphi-c_1\right)\\
{}& c_1,c_2= const\mbox{,  }
\end{split}
\end{equation}
We assume that $p_1$, $p_2$ are limited and rewrite \eqref{ur11}.
The right-hand part of this equality is clearly a limited function at $\lambda\neq 0$, therefore $p_1$, $p_2$ are also limited.

According to \eqref{ur25}, the body moves in the horizontal direction with a mean
velocity of $-\frac\mu\lambda$. This result was obtained in \cite{new036} for $\chi=\zeta=0$.

{\bf Acknowledgement} We thank S.\,M.\,Ramodanov for useful
discussions. This work was supported by the program ``State Support for Leading
Scientific Schools'' (136.2003.1), the Russian Foundation for Basic Research (04-05-64367),
the U.S. Civilian Research and Development Foundation (RU-M1-2583-MO-04)
and the INTAS (04-80-7297).

%\figure {04.eps}

\begin{figure}[ht!]
$$
\includegraphics{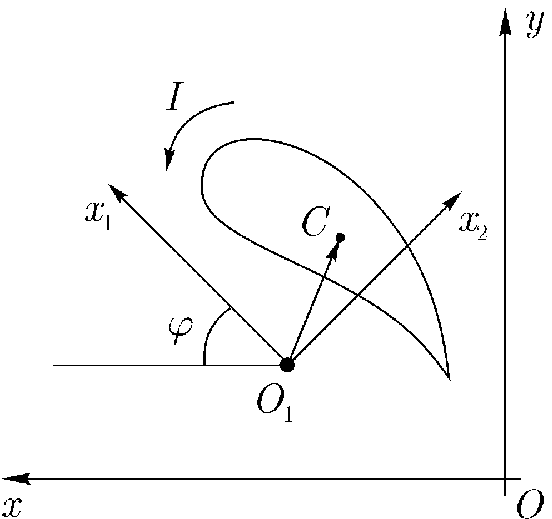}
$$
\caption{}
\end{figure}

\begin{figure}[ht!]
$$
\includegraphics{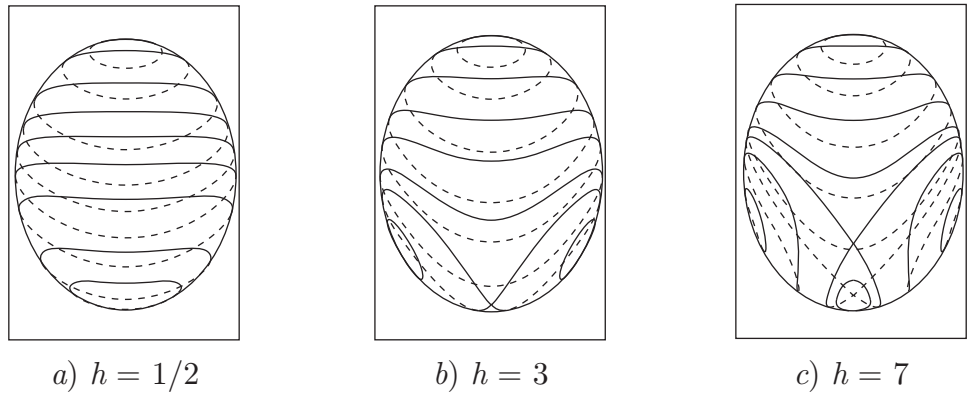}
$$
\caption{Trajectories of the system in Chaplygin's case in the space of variables
$p_1$, $p_2$, $M$. The parameter values are:
$\lambda = 1\mbox{,  } b = 2\mbox{,  } a_1 = 1\mbox{,  } a_2 = \frac{1}{2}$, $\chi = \zeta = 0$.}
\end{figure}

\begin{figure}[ht!]
$$
\psfrag{a}{{\small $h = \frac{1}{2b} c^2$}}
\psfrag{a1}{{\small $\frac{b\lambda^2}{2 a_2^2}$}}
\psfrag{a2}{{\small $\frac{b\lambda^2}{2 a_1^2}$}}
\psfrag{b1}{{\small $\frac{\lambda b}{a_1}$}}
\psfrag{b2}{{\small $\frac{\lambda b}{a_2}$}}
\psfrag{h1}{{\small $h = \frac{\lambda}{a_2}c - \frac{1}{2}\frac{\lambda^2}{a_2^2}b$}}
\psfrag{h2}{{\small $h = \frac{\lambda}{a_1}c - \frac{1}{2}\frac{\lambda^2}{a_1^2}b$}}
\psfrag{a0}{{\small $a_1 > a_2$}}
\psfrag{h}{{\small $h$}}
\psfrag{c}{{\small $c$}}
\includegraphics{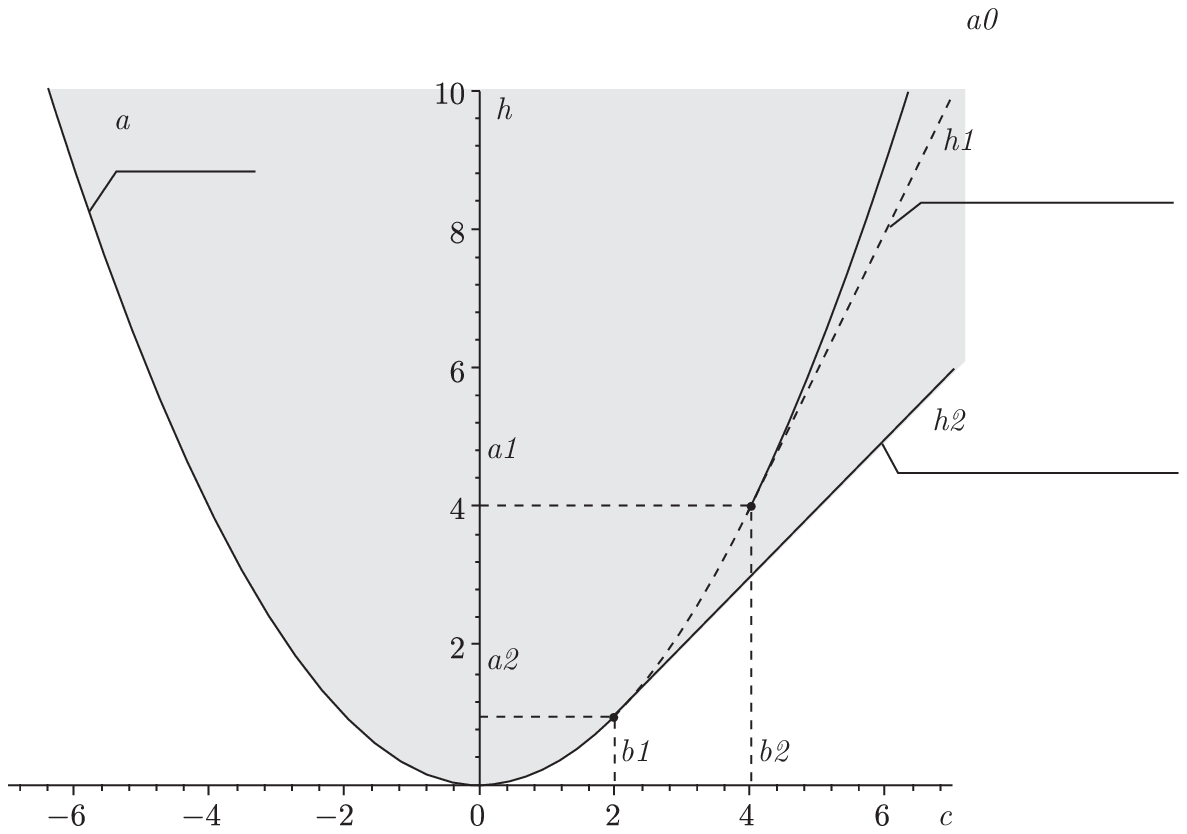}
$$
\caption{Bifurcation diagram for Chaplygin's case at $a_1 > a_2$. The
domain of physically admissible values of integrals $c$, $h$ is shown in
gray. Unstable permanent rotations are shown by dashed lines. Parameter
values are: $\lambda = 1,\ b =2,\ a_1 = 1,\ a_2 = \frac{1}{2}.$}
\end{figure}

\begin{figure}[ht!]
$$
\includegraphics{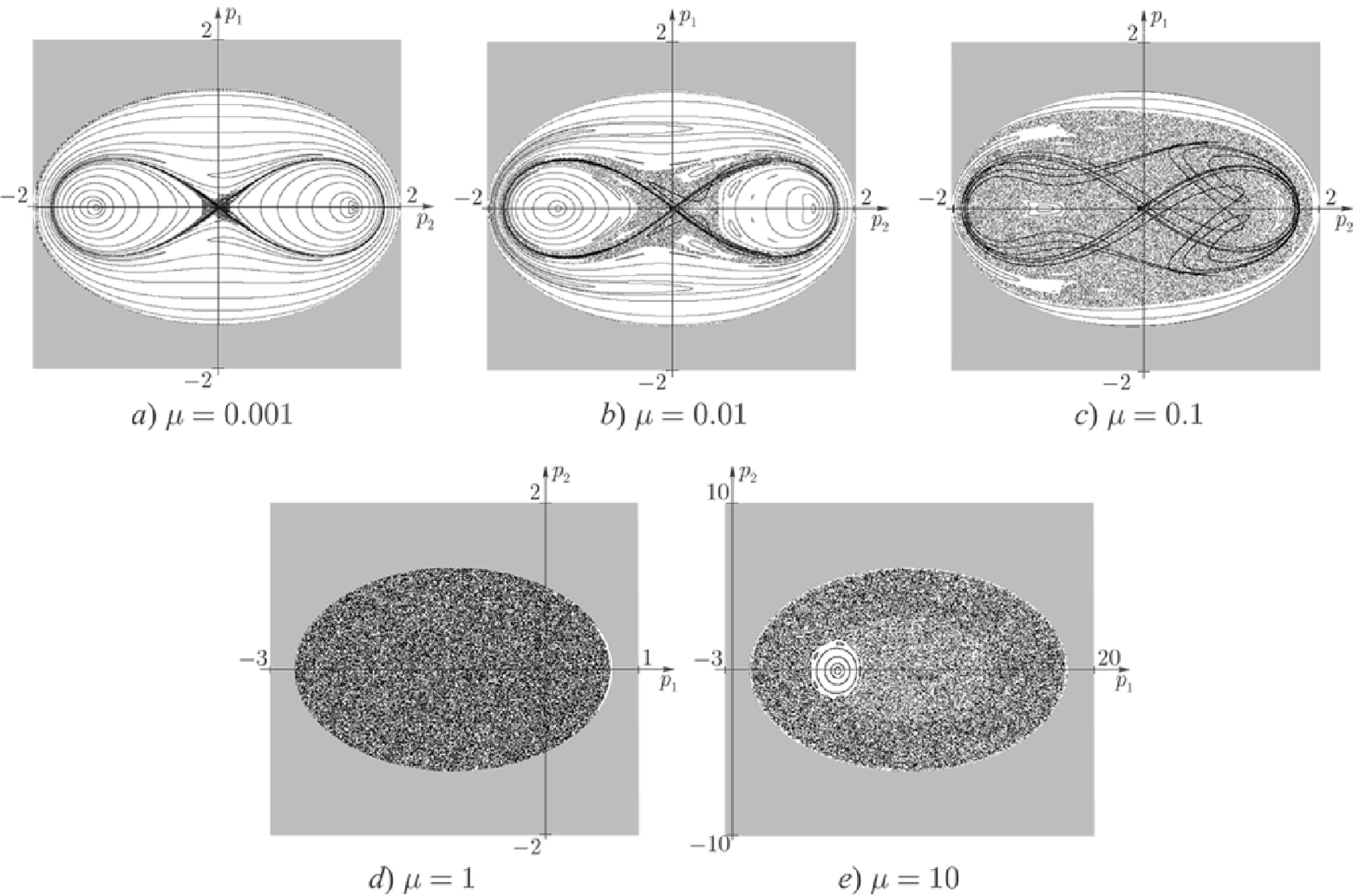}
$$
\caption{Transition to chaos in the Chaplygin system with an increase
in $\mu$. The first three figures show the trajectories and one pair of
separatrices, while, the separatrices are not shown in the latter two
figures because the respective fixed point becomes stable at given
parameter values. (The domain of variables where the motion is impossible
at the given parameter values is shown in gray. The scale of the last
figure differs from that of the others.)}
\end{figure}


\begin{thebibliography}{990}

\bibitem{new015}
Belmonte A., Eisenberg H., Moses E. {\it From flutter to tumble}\/: {\it
inertial drag and Froude similarity in falling paper}, Phys. Rev. Lett.,
1998, Vol.\,81, p. 345--348.

\bibitem{bor10}
Borisov A.\,V., Mamaev I.\,S. {\it Rigid body dynamics}.~--- Izhevsk: SPC
``Regular and chaotic dynamics'', 2001, 384 pp.

\bibitem{bor04}
Chaplygin S.\,A. {\it On heavy body falling in an incompressible fluid} /\!/ Complete Works, Leningrad: Izd. Akad. Nauk SSSR, 1933, Vol.\,1,
p. 133--150.

\bibitem{bor01}
Chaplygin S.\,A. {\it On the effect of a plane-parallel air flow on a
cylindrical wing moving in it} /\!/ Complete Works, Leningrad:
Izd. Akad. Nauk SSSR, 1933, Vol.\,3, p. 3--64.

\bibitem{bor07}
Deryabin M.\,V. {\it On asymptotics of the solution of Chaplygin equation},
Reg. \& Chaot. Dyn., 1998, Vol.\,3, No. 1, p. 93--97.

\bibitem{bor08}
Deryabin M.\,V., Kozlov V.\,V. {\it On the effect of abruptly rising of a
heavy solid body in a fluid}, Izv. RAN, Mekh. tv. tela, 2002, No. 1,
p. 68--74.

\bibitem{new032}
Feild S.\,B., Klaus M., Moore M.\,G., Nori F. {\it Chaotic dynamics of
falling disks}, Nature, 1997, Vol.\,388, p. 252--254.

\bibitem{new006}
Goryachev D.\,N. {\it On the motion of a heavy rigid body in a fluid},
Izv. Imper. Ob-va, Mosk. Imper. Univ., 1893, Vol. 78, No. 2, p. 59--61.

\bibitem{bor06}
Kozlov V.\,V. {\it On a heavy rigid body falling in an ideal fluid},
Izv. Akad. Nauk SSSR, Mekh. tv. tela, 1989, No. 5, p. 10--17.


\bibitem{bor09}
Kozlov V.\,V. {\it On the stability of equilibrium positions in a
non-stationary force field}, PMM, 1991, Vol.\,55, No. 1, p. 12--19.

\bibitem{bor10_2}
Kozlov V.\,V. {\it On the problem of a heavy rigid body falling in a
resistant medium}, Vest. MGU, ser. mat. mekh., 1990, No. 1, p. 79--86.

\bibitem{bor11}
Kozlov V.\,V. {\it On a heavy cylindrical body falling in a fluid}, Izv.
RAN, Mekh. tv. tela, 1993, No. 4, p. 113--117.

\bibitem{bor12}
Kozlov V.\,V. {\it Symmetry, Topology and Resonances in Hamiltonian Mechanics.}~--- Springer-Verlag, 1996.

\bibitem{bor13}
Lamb H. {\em Hydrodynamics}.~--- OGIZ, Gostekhizdat., 1947.
Translated from eng.: Lamb H. {\em Hydrodynamics}, Ed. 6th.~--- N. Y. Dover Publ., 1945.

\bibitem{new036}
Mahadevan L., Ryu N.\,S., Samuel A.\,D.\,T. {\it Tumbling cards}, Phys.
Fluids, 1999, Vol. 11, p. 1--3.

\bibitem{new033}
Maxwell J.\,C. {\it On a particular case of the descent of a heavy body in
a resisting medium}, Camb. Dublin Math. J., 1853, Vol. 9, p. 115--118.

\bibitem{new035}
Pesavento U., Wang Z.\,J. {\it Falling Paper}\/: {\it Navier--Stokes
solutions, model of fluid forces, and center of mass elevation}, Phys.
Rev. Lett., 2004, Vol. 93, No. 14, p. 144--501.

\bibitem{new037}
Ramodanov S.\,M. {\it Asymptotic solutions of the Chaplygin equations}, Vestn. MGU, Ser.
mat. mekh., 1995, No. 3, p. 93--97.

\bibitem{new038}
Ramodanov S.\,M. {\it The effect of circulation on the
fall of a heavy rigid body}, Vestn. MGU, Ser. mat. mekh., 1996, No. 5, p.
19--24.

\bibitem{bor05}
Steklov V.\,A. {\em On the motion of a rigid body in a fluid}.~--- Kharkov, 1893,
234 pp.

\bibitem{new016}
Tanabe Y., Kaneko K. {\it Behavior of a falling paper}, Phys. Rev. Lett.,
1994, Vol.\,73, No. 10, p. 1372--1377.

\bibitem{bor02}
Zhukovski N.\,E. {\it On light elongated bodies that fall in the air while
rotating around the longitudinal axis: I} /\!/ Complete Works,
Moscow--Leningrad: Glav. red. aviats. lit., Vol.\,5, 1937. p. 72--80.

\bibitem{bor03}
Zhukovski N.\,E. {\em On birds' hovering} /\!/ Complete Works,
Moscow-Leningrad: Glav. red. aviats. lit., 1973, Vol.\,5, p. 7--35.

\end{thebibliography}
\end{document}